\documentclass{article}
\usepackage{amsmath}
\usepackage{graphicx}

\begin{document}

\textbf{Lorentz transformations of the electric and magnetic fields}

\textbf{according to Minkowski\bigskip \medskip }

Tomislav Ivezi\'{c}

Ru\mbox
{\it{d}\hspace{-.15em}\rule[1.25ex]{.2em}{.04ex}\hspace{-.05em}}er Bo\v{s}%
kovi\'{c} Institute, P.O.B. 180, 10002 Zagreb, Croatia

E-mail: ivezic@irb.hr\textit{\bigskip \bigskip }

\noindent The usual transformations (UT) of the 3-vectors $\mathbf{E}$ and $%
\mathbf{B}$ that are found by Lorentz, Poincar\'{e} and independently by
Einstein in 1905. are generally considered to be the Lorentz transformations
(LT) of $\mathbf{E}$ and $\mathbf{B}$. According to the UT $\mathbf{E}$ in
one frame is `seen'\ as $\mathbf{E}^{\prime }$ and $\mathbf{B}^{\prime }$ in
a relatively moving frame. In Minkowski's last paper, in 1908. in section
11.6, he defined the vectors (with four components) of the electric $\Phi $
and magnetic $\Psi $ fields and discovered that, e.g., $\Phi $ correctly
transforms by the LT again to $\Phi ^{\prime }$. His correct LT are
reinvented in, e.g., [11] ([11] Ivezi\'{c} T 2005 \textit{Found. Phys.}
\textit{Lett. }\textbf{18} 301). In this paper we show the essential
similarity and some differences between Minkowski's relations in section
11.6 and the results obtained in [11]. The low-velocity limit of the UT and
the LT is briefly examined. A short discussion of the comparison with the
Trouton-Noble experiment is presented.\bigskip \medskip

\bigskip \bigskip

\noindent \textbf{1. Introduction\bigskip }

\noindent It is generally accepted by physics community that there is an
agreement between the classical electromagnetism and the special relativity
as it is formulated in [1]. Both, in the prerelativistic physics and in the
special relativity the electric and magnetic fields are represented by the
3-vectors $\mathbf{E(r,}t\mathbf{)}$ and $\mathbf{B(r,}t\mathbf{)}$. The
notation in this Introduction is as in \ [2], i.e. $\mathbf{E}$ and $\mathbf{%
B}$ are called 3-vectors and they are designated in boldface type. (In the
rest of this paper the notation is changed according to the discussion that
is presented immediately below the Introduction.) In the usual covariant
approaches the field-strength tensor $F^{\alpha \beta }$ (only components)
is introduced and defined in terms of the 4-vector potential $A^{\mu }$ (the
Greek indices run from 0 to 3), equation (11.136) in [2]. The six components
of $F^{\alpha \beta }$ are defined to be six components of the 3-vectors $%
\mathbf{E}$ and $\mathbf{B}$, equation (11.137) in [2]. It is worth noting
that such an identification of the components of $\mathbf{E}$ and $\mathbf{B}
$ with the components of $F^{\alpha \beta }$ is synchronization dependent as
explicitly shown in [3]. This is also discussed in [4]. The 3-vector $%
\mathbf{E}$ is constructed as $\mathbf{E=}F^{10}\mathbf{i}+F^{20}\mathbf{j}%
+F^{30}\mathbf{k}$. The usual transformations (UT) of the components of $%
\mathbf{E}$ and $\mathbf{B}$ are derived assuming that they transform under
the Lorentz transformations (LT) (boosts) as the components of $F^{\alpha
\beta }$ transform, equation (11.148) in [2]. Then $\mathbf{E}^{\prime }$
and $\mathbf{B}^{\prime }$ are constructed in the inertial frame of
reference $S^{\prime }$ in the same way as in $S$, i.e. multiplying the
components $E_{x,y,z}^{\prime }$ and $B_{x,y,z}^{\prime }$ by the unit
3-vectors $\mathbf{i}^{\prime }$, $\mathbf{j}^{\prime }$, $\mathbf{k}%
^{\prime }$. This yields the UT of $\mathbf{E}$ and $\mathbf{B}$\textbf{, }%
equation (11.149) in [2], i.e. equations (\ref{JC1}) and (\ref{JB}) here.
Observe that there are no LT, or any other transformations, that transform
the unit 3-vectors $\mathbf{i}$, $\mathbf{j}$, $\mathbf{k}$ into the unit
3-vectors $\mathbf{i}^{\prime }$, $\mathbf{j}^{\prime }$, $\mathbf{k}%
^{\prime }$. It is seen from equations (11.148) and (11.149) in [2], i.e.
from equation (\ref{JC1}) here, that the transformed $\mathbf{E}^{\prime }$
is expressed by the mixture of the 3-vectors $\mathbf{E}$ and $\mathbf{B}$,
and similarly for $\mathbf{B}^{\prime }$, as seen from (\ref{JB}). The UT,
equations (11.148) and (11.149) in [2], are always considered to be the
relativistically correct LT (boosts) of $\mathbf{E}$ and $\mathbf{B}$. They
are first derived by Lorentz [5] and Poincar\'{e} [6] (see also two
fundamental Poincar\'{e}'s papers with notes by Logunov [7]) and
independently by Einstein [1]. Einstein's derivation of the UT of $\mathbf{E}
$ and $\mathbf{B}$ is objected and discussed in detail in section 5.3. in
[3].

However, in 1908., Minkowski, in section 11.6 in [8], defined the electric
and magnetic fields on the four-dimensional (4D) spacetime, equation (\ref%
{M1}) here, and gave a general form of the mathematically correct LT of such
4-vector fields, equation (\ref{M3}) here. According to equation (\ref{M3})
\emph{a 4-vector of the electric field transforms by the LT as any other
4-vector transforms; i.e. it transforms again to the 4-vector of the
electric field.} \emph{There is no mixing with components of magnetic field.}
His mathematically correct LT of fields remained almost completely unknown.
They are not mentioned even in the recent publications in - Annalen der
Physik, Special Topic Issue 9-10/2008: The Minkowski spacetime of special
relativity - 100 years after its discovery. Perhaps, one of the reason for
such systematic neglect of that important Minkowski's contribution was that
Minkowski himself never applied these transformations of the 4-vector
fields. In all other parts of [8] he dealt with the usual 3-vectors $\mathbf{%
E}$ and $\mathbf{B}$. In [9], for the first time, it was observed the
importance and the relativistic correctness of section 11.6 in [8] and also
the apparent similarity of the mentioned Minkowski's results with the recent
results obtained in [10-13]. It is proved in [10-13] that the LT always
transform an algebraic object defined on the 4D spacetime that represent the
electric field only to the electric field, and similarly for the magnetic
field, as in (\ref{LM3})-(\ref{LTE2}).

In this paper we shall investigate the similarity and some differences
between Minkowski's results, section 11.6 in [8], and the results from
[10-13]. It is shown that Minkowski's relations (\ref{M1}), (\ref{M2}) and (%
\ref{M3}) correspond to the relations (\ref{E1}), (\ref{E2}) and (\ref{LM3})
respectively, which are obtained in, e.g., [11]. The relations (\ref{LTE1})
and (\ref{LTE2}), i.e. the explicit forms of the LT of the electric field,
were not discovered by Minkowski. They are derived transforming both the $F$
field and the observer $\gamma _{0}$ and they are reported in [10-13]. When
only $F$ is transformed by the LT, but not the observer $\gamma _{0}$, then
the transformation of $E=F\cdot \gamma _{0}$ is given by equation (\ref{J1}%
). It is shown that the components of the transformed $E_{F}^{\prime }$ are
nothing else than the components that are obtained by the UT, equation
(11.148) in [2]. This result undoubtedly reveals that the UT of the
3-vectors $\mathbf{E}$ and $\mathbf{B}$, equations (\ref{JC1}) and (\ref{JB}%
), differ from the correct LT (\ref{LM3})-(\ref{LTE2}). In the last part of
this paper the low-velocity limit of the UT and the LT is briefly discussed.
Also, a short discussion of the comparison with the Trouton-Noble experiment
is presented.\bigskip \bigskip

\noindent \textbf{2. The Lorentz transformations. Both }$F$ \textbf{and the
observer }

\textbf{are transformed\bigskip }

\noindent First, let us expose an important result regarding the usual
formulation of electromagnetism (as in [2]), which is presented in [9]. This
is also mentioned in [4]. It is explained in [9] that an individual vector
has no dimension; the dimension is associated with the vector space and with
the manifold where this vector is tangent. Hence, what is essential for the
number of components of a vector field is the number of variables on which
that vector field depends, i.e., the dimension of its domain. This means
that the usual time-dependent $\mathbf{E(r,}t\mathbf{)}$, $\mathbf{B(r,}t%
\mathbf{)}$ cannot be the 3-vectors, since they are defined on the
spacetime. That fact determines that such vector fields, when represented in
some basis, have to have four components (some of them can be zero).
Therefore, from now on, we shall use the term `vector' for the correctly
defined geometric quantities, which are defined on the spacetime. (In
section 1, they are called, as in [2], the 4-vectors.) However, an incorrect
expression, the 3-vector or the 3D vector, will still remain for the usual $%
\mathbf{E(r,}t\mathbf{)}$, $\mathbf{B(r,}t\mathbf{)}$ from [2], see
equations (\ref{JC1}) and (\ref{JB}).

Our consideration will be in the geometric algebra formalism. A brief review
of the geometric algebra is given here, but for more detail see [14]. The
geometric product (it is written by simply juxtaposing multivectors $AB$) of
a grade-$r$ multivector $A_{r}$ with a grade-$s$ multivector $B_{s}$
decomposes into $A_{r}B_{s}=\left\langle AB\right\rangle _{\
r+s}+\left\langle AB\right\rangle _{\ r+s-2}...+\left\langle AB\right\rangle
_{\ \left\vert r-s\right\vert }$. The inner and outer (or exterior) products
are the lowest-grade and the highest-grade terms respectively of the above
series; $A_{r}\cdot B_{s}\equiv \left\langle AB\right\rangle _{\ \left\vert
r-s\right\vert }$ and $A_{r}\wedge B_{s}\equiv \left\langle AB\right\rangle
_{\ r+s}$. For vectors $a$ and $b$ we have: $ab=a\cdot b+a\wedge b$, where $%
a\cdot b\equiv (1/2)(ab+ba)$, $a\wedge b\equiv (1/2)(ab-ba)$. The generators
of the spacetime algebra are four basis vectors $\left\{ \gamma _{\mu
}\right\} ,\mu =0...3,$ satisfying $\gamma _{\mu }\cdot \gamma _{\nu }=\eta
_{\mu \nu }=diag(+---)$. This basis, the standard basis, is a right-handed
orthonormal frame of vectors in the Minkowski spacetime $M^{4}$ with $\gamma
_{0}$ in the forward light cone, $\gamma _{0}^{2}=1$ and $\gamma _{k}^{2}=-1$
($k=1,2,3$). The standard basis $\left\{ \gamma _{\mu }\right\} $
corresponds to Einstein's system of coordinates in which the Einstein
synchronization [1] of distant clocks and Cartesian space coordinates $x^{i}$
are used in the chosen inertial frame of reference. The unit pseudoscalar $I$
from equations (\ref{E1}), (\ref{E2}) and (\ref{J1}) is defined
algebraically without introducing any reference frame, as in section 1.2. in
the second reference in [14]. When $I$ is represented in the $\left\{ \gamma
_{\mu }\right\} $ basis it becomes $I=\gamma _{0}\wedge \gamma _{1}\wedge
\gamma _{2}\wedge \gamma _{3}$.

In equation (23) in [11], the electric and magnetic fields are represented
by vectors $E(x)$ and $B(x)$.\ The electromagnetic field is represented by
the bivector $F=F(x)$ and $v$ denotes the velocity vector of a family of
observers who measures $E$ and $B$ fields. Then
\begin{equation}
E=(1/c)F\cdot v,\quad B=-(1/c^{2})I(F\wedge v).  \label{E1}
\end{equation}%
Note that $E$ and $B$ in (\ref{E1}) depend not only on $F$ but on $v$ as
well.

These relations correspond to Minkowski's relations from section 11.6
\begin{equation}
\Phi =-wF,\quad \Psi =iwf^{\ast }.  \label{M1}
\end{equation}%
(In the vacuum $f=F$ and one could write the second equation in (\ref{M1})
as $\Psi =iwF^{\ast }$, where $F^{\ast }$ is the dual field-strength tensor,
$^{\ast }F^{\alpha \beta }=(1/2)\varepsilon ^{\alpha \beta \gamma \delta
}F_{\gamma \delta }$.)

Observe that (\ref{E1}) are coordinate-free relations, which hold for any
observer. When geometric quantities from (\ref{E1}) are represented in some
basis then they contain both components and basis vectors. In contrast to
it, Minkowski considered that $w$, $\Phi $ and $\Psi $ are $1\times 4$
matrices and $F$ is a $4\times 4$ matrix. Their components are implicitly
determined in the standard basis.

In equation (23) in [11], the decomposition of $F$ in terms of vectors $E$, $%
B$ and $v$ is given as
\begin{equation}
F=(1/c)E\wedge v+(IB)\cdot v,  \label{E2}
\end{equation}%
where, from (\ref{E2}) and (\ref{E1}), it holds that $E\cdot v=B\cdot v=0$.

The relation that corresponds to (\ref{E2}) is equation (55) in [8]%
\begin{equation}
F=[w,\Phi ]+i\mu \lbrack w,\Psi ].  \label{M2}
\end{equation}

In section 11.6 in [8], the next paragraph below equation (44), Minkowski
described how $w$ and $F$ separately transform under the LT $A$ (the matrix
of the LT is denoted as $A$ in [8]) and then how the product $wF$
transforms. Thus, he wrote $w^{\prime }=wA$ for the LT of the velocity
vector $w$ and $F^{\prime }=A^{-1}FA$ for the LT of the field-strength
tensor. Then the mathematically correct LT of $wF$ are
\begin{equation}
\Phi =wF\longrightarrow \Phi ^{\prime }=wAA^{-1}FA=(wF)A=\Phi A,  \label{M3}
\end{equation}%
which means that under the LT both terms, the velocity $w$ and $F$ are
transformed and their product transforms as any other vector (i.e., in [8],
an $1\times 4$ matrix) transforms. The most important thing is that \emph{%
the electric field vector }$\Phi $ \emph{transforms by the LT again to the
electric field vector }$\Phi ^{\prime }$\emph{; there is no mixing with the
magnetic field} $\Psi $.

These correct LT of the electric and magnetic fields are reinvented in
[10-13]. Let us choose the frame in which the observers who measure $E$ and $%
B$ from (\ref{E1}) are at rest. For them $v=c\gamma _{0}$. In the geometric
algebra the LT are described with rotors $R$, $R\widetilde{R}=1$, where the
reverse $\widetilde{R}$ is defined by the operation of reversion according
to which $\widetilde{AB}=\widetilde{B}\widetilde{A}$, for any multivectors $%
A $ and $B$, $\widetilde{a}=a$, for any vector $a$, and it reverses the
order of vectors in any given expression. For boosts in arbitrary direction
the rotor $R$ is given by equation (8) in [11,13] as
\begin{equation}
R=(1+\gamma +\gamma \gamma _{0}\beta )/(2(1+\gamma ))^{1/2},  \label{LTR}
\end{equation}%
where $\gamma =(1-\beta ^{2})^{-1/2}$, the vector $\beta $ is $\beta =\beta
n $, $\beta $ on the r.h.s. of that equation is the scalar velocity in units
of $c$ and $n$ is not the basis vector but any unit space-like vector
orthogonal to $\gamma _{0}$. Then, \emph{any multivector} $M$ \emph{%
transforms by active LT in the same way}, i.e. as in equation (9) in [13],
\begin{equation}
M\rightarrow M^{\prime }=RM\widetilde{R}.  \label{RM}
\end{equation}%
Hence, vector $E$ transforms by the LT $R$ as $E\longrightarrow E^{\prime
}=RE\widetilde{R}$. When $v=c\gamma _{0}$ is taken in (\ref{E1}) then $E$
becomes $E=F\cdot \gamma _{0}$ and it transforms under the LT in the same
manner as in (\ref{M3}), i.e., that both $F$ and $v$ are transformed by the
LT $R$ as
\begin{equation}
E=F\cdot \gamma _{0}\longrightarrow E^{\prime }=(RF\widetilde{R})\cdot
(R\gamma _{0}\widetilde{R})=R(F\cdot \gamma _{0})\widetilde{R}.  \label{LM3}
\end{equation}%
These correct LT give that

\begin{equation}
E^{\prime }=E+\gamma (E\cdot \beta )\{\gamma _{0}-(\gamma /(1+\gamma ))\beta
\}.  \label{LTE1}
\end{equation}%
In the same way every vector transforms, i.e., the vector $B$ as well. For
boosts in the direction $\gamma _{1}$ one has to take that $\beta =\beta
\gamma _{1}$ (on the l.h.s. is vector $\beta $ and on the r.h.s. $\beta $ is
a scalar) in the above expression for the rotor $R$ (all in the standard
basis). Hence, in the $\left\{ \gamma _{\mu }\right\} $ basis and when $%
\beta =\beta \gamma _{1}$ equation (\ref{LTE1}) becomes
\begin{equation}
E^{\prime \nu }\gamma _{\nu }=-\beta \gamma E^{1}\gamma _{0}+\gamma
E^{1}\gamma _{1}+E^{2}\gamma _{2}+E^{3}\gamma _{3},  \label{LTE2}
\end{equation}%
what is equation (9) in [11]. The same components would be obtained for $%
\Phi ^{\prime }=\Phi A$ in Minkowski's relation (\ref{M3}) when the
components of $w$ are $(0,0,0,ic)$ in his notation, which corresponds to $%
v=c\gamma _{0}$ in our formulation.

As already stated, equations (\ref{LTE1}) and (\ref{LTE2}) are derived in
[10-13]. Minkowski wrote (\ref{M1}), (\ref{M2}) and (\ref{M3}) in section
11.6 in [8], but in the rest of [8] he exclusively dealt with the usual
3-vectors $\mathbf{E}$ and $\mathbf{B}$ and not with correctly defined
vectors $\Phi $ and $\Psi $.

If one represents the relation $E=(1/c)F\cdot v$ from (\ref{E1}) in the
standard basis $\left\{ \gamma _{\mu }\right\} $ then $E=E^{\mu }\gamma
_{\mu }$, where $E^{\mu }=F^{\mu \nu }v_{\nu }$ (e.g., $%
E^{1}=F^{10}v_{0}+F^{12}v_{2}+F^{13}v_{3}$ and $%
E^{0}=F^{01}v_{1}+F^{02}v_{2}+F^{03}v_{3}$). These relations for components
exactly correspond to Minkowski's expressions for the relation $\Phi =-wF$
in components, when the components $\Phi _{1}$, .., $\Phi _{4}$ are
expressed in terms of the components $w_{1}$, .., $w_{4}$ and the components
$F_{hk}$ (in [8] $h$, $k=1,2,3,4$ and $h=4$ denotes imaginary time
component). Thus, e.g., $\Phi _{1}=w_{4}F_{14}+w_{2}F_{12}+w_{3}F_{13}$ and $%
\Phi _{4}=w_{1}F_{41}+w_{2}F_{42}+w_{3}F_{43}$. Minkowski was a very good
mathematician and he completely understood that \emph{mathematically}
correct Lorentz transformations of fields are those of his $\Phi $, (\ref{M3}%
). But, probably, due to the generally accepted belief and the authorities
in physics (Maxwell, Lorentz, Einstein, ..), he also believed that physical
quantities are the usual 3-vectors $\mathbf{E}$, $\mathbf{B}$ and $\mathbf{D}
$, $\mathbf{H}$. Therefore he expressed in equations (47), (48) and (51),
(52) in [8] the components of his mathematically and \emph{physically}
correct fields (in my opinion) $\Phi $ and $\Psi $ in terms of the usual
3-vectors $\mathbf{E}$, $\mathbf{B}$ and $\mathbf{D}$, $\mathbf{H}$. He
wrote for equation (47) in [8] that the first three components $\Phi _{1}$, $%
\Phi _{2}$, $\Phi _{3}$ are the components of the 3-vector $(\mathbf{E}+%
\mathbf{w}\times \mathbf{M})/(1-w^{2})^{-1/2}$, whereas $\Phi _{4}=i(\mathbf{%
w}\cdot \mathbf{E})/(1-w^{2})^{-1/2}$ (his $\mathbf{M}$ is our $\mathbf{B}$%
). He called $\Phi $ - the electric field at rest - and similarly $\Psi $ -
the magnetic field at rest - because it follows from equations (47) and (48)
that for his $w=(0,0,0,ic)$ the temporal component $\Phi _{4}=0$ and the
spatial components $\Phi _{1}$, $\Phi _{2}$, $\Phi _{3}$ are the same as the
components of the usual electric field 3-vector $\mathbf{E}$ and similarly
for $\Psi $. So, he believed that only when $w=(0,0,0,ic)$ his fields $\Phi $
and $\Psi $ are the electric and magnetic fields since then they coincide
with `physical' $\mathbf{E}$ and $\mathbf{B}$. However, regardless of that
problem with physical interpretation, or, better to say, because of that
problem, Minkowski's section 11.6 is very important for all
physicists.\bigskip \bigskip

\noindent \textbf{3. The usual transformations. Only }$F$ \textbf{is
transformed but not the observer\bigskip }

\noindent Now, let us see what will be obtained if in the transformation of $%
E=F\cdot \gamma _{0}$ only $F$ is transformed by the LT $R$, but not the
velocity of the observer $v=c\gamma _{0}$. Of course, it will not be the LT
of $E=F\cdot \gamma _{0}$, since they are given by (\ref{LM3}). Thus
\begin{equation}
E=F\cdot \gamma _{0}\longrightarrow E_{F}^{\prime }=(RF\widetilde{R})\cdot
\gamma _{0}.  \label{EP1}
\end{equation}%
This yields that

\begin{equation}
E_{F}^{\prime }=\gamma \{E+(\beta \wedge \gamma _{0}\wedge cB)I\}+(\gamma
^{2}/(1+\gamma ))\beta (\beta \cdot E),  \label{J1}
\end{equation}%
which, in the standard basis and when $\beta =\beta \gamma _{1}$, becomes

\begin{equation}
E_{F}^{\prime \nu }\gamma _{\nu }=E^{1}\gamma _{1}+\gamma (E^{2}-c\beta
B^{3})\gamma _{2}+\gamma (E^{3}+c\beta B^{2})\gamma _{3}.  \label{J2}
\end{equation}%
The transformation (\ref{J1}) can be compared with the UT for the 3-vector $%
\mathbf{E}$ that are given, e.g. by equation (11.149) in [2],\ i.e. with
\begin{equation}
\mathbf{E}^{\prime }=\gamma (\mathbf{E}+\mathbf{\beta \times }c\mathbf{B)-}%
(\gamma ^{2}/(1+\gamma ))\mathbf{\beta (\beta \cdot E)}  \label{JC1}
\end{equation}%
and equation (\ref{J2}) with equation (11.148) in [2]. In (\ref{JC1}) $%
\mathbf{E}^{\prime }$, $\mathbf{E}$, $\mathbf{\beta }$ and $\mathbf{B}$ are
all 3-vectors. It is visible from the comparison of equation (\ref{J2}) with
equation (11.148) in [2] that the transformations of components (taken in
the standard basis) of $E_{F}^{\prime }$ are exactly the same as the
transformations of $E_{x,y,z}$ from equation (11.148) in [2]. The UT for $%
\mathbf{B}$ are given by the second equation in equation (11.149) in [2],
\begin{equation}
\mathbf{B}^{\prime }=\gamma (\mathbf{B}-(1/c)\mathbf{\beta \times E)-}%
(\gamma ^{2}/(1+\gamma ))\mathbf{\beta (\beta \cdot B)}  \label{JB}
\end{equation}%
The transformations (\ref{EP1}) and (\ref{J2}) are first discussed in detail
in [10-13] and compared with the UT (11.148) and (11.149) from [2], whereas
the general form of $E_{F}^{\prime }$, equation (\ref{J1}), is first given
in [9].

Here, it is at place to point out an important difference between the LT and
the UT. If instead of the active LT we consider the passive LT then, e.g.
the vector $E=E^{\nu }\gamma _{\nu }=E^{\prime \nu }\gamma _{\nu }^{\prime }$
will remain unchanged, because the components $E^{\nu }$ transform by the LT
and the basis vectors $\gamma _{\nu }$ by the inverse LT leaving the whole $%
E $ invariant under the passive LT. Of course, the same holds for all bases
including those with nonstandard synchronizations, as shown, e.g., in [3].
\emph{This invariance of} $E$ \emph{under the LT means that the electric
field} $E$ \emph{is the same physical quantity for all relatively moving
observers.} It is not so with the 3-vector $\mathbf{E}$ and its UT. Namely, $%
\mathbf{E=}E_{x}\mathbf{i}+E_{y}\mathbf{j}+E_{z}\mathbf{k}$ \ is completely
different than $\mathbf{E}^{\prime }$ from (\ref{JC1}), see the discussion
in section 1. This means that although $\mathbf{E}$ and $\mathbf{E}^{\prime
} $ are measured by different observers \emph{they are not the same quantity
for such relatively moving observers. }The observers are not looking at the
same physical object, here the electric field vector, but at two different
objects. Every observer makes measurement of its own 3-vector field, $%
\mathbf{E}$ and $\mathbf{E}^{\prime }$, and such measurements are not
related by the LT. Different relatively moving inertial 4D observers can
compare only 4D quantities, here $E^{\nu }\gamma _{\nu }$ and $E^{\prime \nu
}\gamma _{\nu }^{\prime }$, because they are connected by the LT. The
experimentalists have to measure \emph{all components} of 4D quantities,
here of $E$, in both frames $S^{\prime }$ and $S$. The observers in $%
S^{\prime }$ and $S$ are able to compare only such complete set of data
which corresponds to the \emph{same} 4D geometric quantity.\bigskip \bigskip

\noindent \textbf{4. The low-velocity limit of the UT and the LT. A short
discussion }

\textbf{of the comparison with the Trouton-Noble experiment} \textbf{%
\bigskip }

\noindent When\textbf{\ }the low-velocity limit $\beta \ll 1$, or $\gamma
\simeq 1$, is taken in (\ref{JC1}) and in (\ref{JB}), then the following
relations with 3-vectors are obtained $\mathbf{E}^{\prime }=\mathbf{E}+%
\mathbf{\beta \times }c\mathbf{B}$ and $\mathbf{B}^{\prime }=\mathbf{B}-(1/c)%
\mathbf{\beta \times E}$. They are commonly used in literature. However, it
is argued in [15] that these transformations have to be replaced by two
well-defined Galilean limits, the magnetic and electric limits, i.e. with
two sets of low-velocity formulae. These two limits are obtained from the UT
(\ref{JC1}) and (\ref{JB}). In vacuum, the magnetic limit is obtained taking
in the UT that not only $\beta \ll 1$, but $\left\vert \mathbf{E}\right\vert
\ll c\left\vert \mathbf{B}\right\vert $ as well. Hence, the UT in the
magnetic limit are: $\mathbf{E}^{\prime }=\mathbf{E}+\mathbf{\beta \times }c%
\mathbf{B}$ and $\mathbf{B}^{\prime }=\mathbf{B}$. Conversely, the electric
limit is obtained taking in the UT (\ref{JC1}) and (\ref{JB}) that $\beta
\ll 1$ and $\left\vert \mathbf{E}\right\vert \gg c\left\vert \mathbf{B}%
\right\vert $. Hence, the UT in the electric limit are: $\mathbf{E}^{\prime
}=\mathbf{E}$ and $\mathbf{B}^{\prime }=\mathbf{B}-(1/c)\mathbf{\beta \times
E}$. The results from [15] are used, developed and applied to different
problems in a series of papers in [16]. Observe that in all papers in [15,
16] the UT of $\mathbf{E}$ and $\mathbf{B}$ are considered to be the
relativistically correct LT.

However, as shown in [10-13], and also here, the UT are not the LT; the LT
are given by equations (\ref{LTE1}), (\ref{LTE2}) and the same for $%
B^{\prime }$. This means that neither the commonly used set of low-velocity
transformations nor the two mentioned limits are the low-velocity
approximations of the LT. In the UT, equations (\ref{JC1}) and (\ref{JB}),
the components of the electric and magnetic fields are mixed together and
therefore it is possible to compare their moduli and to obtain two different
limits. For the LT (\ref{LTE1}) and (\ref{LTE2}) there is only one
low-velocity approximation, which is simply obtained taking the limit $\beta
\ll 1$, or $\gamma \simeq 1$. In that approximation the LT (\ref{LTE1})
become $E^{\prime }\simeq E+(E\cdot \beta )\gamma _{0}$,$\ $and the same for
the vector $B$. It can be easily shown that to order $0(\beta ^{2})$ this
low-velocity approximation of the vector $E$ is invariant under the passive
LT.

In section VIII. A. in the third paper in [16] (it will be denoted as
MR[16]), the electric limit approximation of the UT is used in a comparison
with the Trouton-Noble experiment. The consideration from MR [16] will be
briefly discussed here. First, the authors show that with the common form of
the UT (\ref{JB}) there is an electric energy associated with the motional
magnetic field, equation (52) in MR[16]. As a consequence, there is the
electrical torque in the ether frame, equation (53) in MR[16], although
there is no torque in the rest frame of the capacitor. Then, they consider
the electric limit approximation of the Poynting theorem, equation (54) in
MR. It is visible from that equation that `.. the energy density is of
electric origin only.' and `..no electric energy associated with the
motional magnetic field can be taken into account within the electric limit,
because it is of order $(v/c)^{2}$ with respect to the static, or
quasistatic, electric limit.' Consequently, it is concluded in MR[16] that
\textbf{`..}the Trouton-Noble experiment does not show any effect in the
electric limit.' and also `Of course, special relativity is needed for
larger velocities, and we must take into account the additional mechanical
torque$^{41}$ due to the length variation to explain the negative result
(that is, no torque).' (Ref. [41] is Pauli's book, Pauli W 1981 \textit{%
Theory of relativity} (New York: Dover), my remark.) Strictly speaking,
these two statements contradict each other. According to the second
statement there is the Trouton-Noble paradox (there is a 3D torque in one
frame but no 3D torque in relatively moving frame) for larger velocities.
According to the first statement, there is not the Trouton-Noble paradox
when the electric limit of the low-velocity approximation is used. Hence,
the principle of relativity is violated for larger velocities but not
violated in the low-velocity approximation. Such a result clearly indicates
that both, the approach with the UT, equations (\ref{JC1}) and (\ref{JB}),
and its two low-velocity limits from [15, 16], are not relativistically
correct. Namely, the principle of relativity has to be satisfied for all
velocities less than velocity of light. Furthermore, Pauli's `resolution' of
the Trouton-Noble paradox by the introduction of the additional mechanical
torque also deals with the 3-vectors and their UT and with the length
contraction and the fictive energy current, von Laue's energy current. It is
explicitly shown in [3], in [17] and [18], that, contrary to the general
belief, the Lorentz contraction, and the time dilatation, [3, 18], have
nothing to do with the LT, i.e., with the special relativity as the theory
of the flat spacetime. Namely, in the 4D spacetime, the Lorentz contraction
is meaningless, because it is not possible to compare two spatial lengths
that are simultaneously determined with respect to relatively moving
observers. Besides, it is unobservable. Moreover, as already objected in
[19], von Laue's energy current is something like the phlogiston or the
ether; it carries energy but it cannot be measured. Thus, contrary to the
assertions from MR[16], the electric limit approximation of the Poynting
theorem, which deals with the 3-vectors $\mathbf{E}$ and $\mathbf{B}$, does
not resolve the Trouton-Noble paradox.

In the recent paper [20] it is argued that the Trouton-Noble paradox is
resolved once the electromagnetic momentum (3D quantity) of the moving
capacitor is properly taken into account. In [20], as in all previous
`explanations,' the 3D quantities $\mathbf{E}$, $\mathbf{B}$, $\mathbf{F}$,
the torque $\mathbf{T}=\mathbf{r}\times \mathbf{F}$, the density of
electromagnetic momentum $\mathbf{g=}\varepsilon _{0}\mathbf{E}\times
\mathbf{B}$, etc., are considered to be physical ones in the 4D spacetime
and their UT are used as that they are the LT.

However, it is shown in [21, 22] that in the geometric approach with 4D
quantities \emph{the 4D torques} will not appear for the moving capacitor if
they do not exist for the stationary capacitor, which means that with 4D
geometric quantities the principle of relativity is naturally satisfied and
there is not the Trouton-Noble paradox. Of course, the same conclusion will
hold in the low-velocity approximation $\beta \ll 1$, or $\gamma \simeq 1$.
Thus, there is no need either for the nonelectromagnetic forces and their
additional torque, as, for example, in Pauli's explanation, or for the
angular electromagnetic field momentum and its rate of change, i.e. its
additional torque, as in [20].

In [21, 22], the torque $N$ (bivector) is defined as $N=r\wedge K$,$\ $where
$r=x_{P}-x_{O}$ and $K$, in this problem, is the Lorentz force $K_{L}$. $r$
is the distance vector that is associated with the lever arm in the
Trouton-Noble experiment, $x_{P}$ and $x_{O}$ are the position vectors
associated with the spatial point of the axis of rotation and the spatial
point of application of the force $K_{L}$. $P$ and $O$ are the events whose
position vectors are $x_{P}$ and $x_{O}$. The Lorentz force $K_{L}$ is $%
K_{L}=(q/c)F\cdot u$ ($u$ is the velocity vector).

Notice an essential difference between the treatment of the Trouton-Noble
experiment from [21, 22] and all usual treatments, e.g., [20] and MR[16]. In
[21, 22] it is dealt with events, position vectors, distance vectors, the
Lorentz force vector, the bivectors $F$ and $N$, etc., which are considered
to be physical quantities that are well-defined on the 4D spacetime. In
contrast to it the usual approaches consider that the spatial points, the
spatial distance $\left\vert \mathbf{r}\right\vert $\textbf{, }the 3-vectors
of\textbf{\ }the Lorentz force $\mathbf{F}_{L}$ and the torque $\mathbf{T}$,
etc., are physical quantities.

Observe that in [21] it is exclusively dealt with $F$ and $N$ without using
their decompositions. In [22], the decomposition of $F$ into vectors $E$, $B$
and $v$, (\ref{E2}) (and (\ref{E1})), and a similar decomposition for $N$,
are employed. The torque $N$ is decomposed into two vectors, the
`space-space'\ torque $N_{s}$ and the `time-space'\ torque $N_{t}$ (they
correspond to $B$ and $E$, respectively, in (\ref{E2}) and (\ref{E1})), and
the unit time-like vector $v/c$, where $v$ is the velocity vector of the
observers who measure $N_{s}$ and $N_{t}$, see equation (2) in [22]. The
bivector $N$ is the primary physical quantity for torques; $N_{s}$ and $%
N_{t} $ are derived from $N$ and $v$ and they are \emph{both} equally good
physical\emph{\ }quantities.

In section 4 in [22], the comparison of the approach with $N_{s}$, $N_{t}$
and $N$ and the usual approach with the 3D torque $\mathbf{T}$ is presented.
Comparing the derivation from [22] of the LT of $N_{s}$ and $N_{t}$ and the
UT of $\mathbf{T}$ with the derivation from this paper of the LT of $E$, (%
\ref{LM3}), (\ref{LTE1}), (\ref{LTE2}) and the UT of $E_{F}$, (\ref{EP1}), (%
\ref{J1}) and (\ref{J2}), i.e., the UT of $\mathbf{E}$ and $\mathbf{B}$,
equations (\ref{JC1}) and (\ref{JB}) respectively, one sees that they are
the same. As already stated $N_{s}$ and $N_{t}$ depend on $N$ and $v$.
Hence, in order to have their LT both $N$ and $v$ have to be transformed by
the LT in the same way as in Minkowski's relation (\ref{M3}), i.e., as in
our relations (\ref{LM3}), (\ref{LTE1}) and (\ref{LTE2}). And, of course, $%
N_{s}$ ($N_{t}$) is transformed by the active LT to $N_{s}^{\prime }$ ($%
N_{t}^{\prime }$). When only $N$ is transformed by the LT and not $v$,
similarly as in (\ref{EP1}), (\ref{J1}) and (\ref{J2}), then the UT of $%
N_{s} $ and $N_{t}$ are obtained and they differ from the correct LT.

In order to test special relativity, e.g., by means of the Trouton-Noble
type experiments, it is not enough, as usually done, to measure three
independent parameters of the 3D rotation, i.e., three independent
components of $N_{s}$, but also one has to measure the other three relevant
variables, i.e., three independent components of $N_{t}$. This essential
difference between the measurements of the 3D quantities and the
relativistically correct 4D geometric quantities is the real cause of the
appearance of different paradoxes in the usual, Einstein's formulation of
special relativity. Such a paradox, which is very similar to the
Trouton-Noble paradox, is Jackson's paradox. It is discussed in detail in
[23]; the second paper is a more pedagogical version of the first one.

Particularly interesting, and potentially very important application of such
decompositions as for $F$ and $N$, is presented in [4]. (In general, any
second rank antisymmetric tensor can be decomposed into two vectors and a
unit time-like vector (the velocity vector/c). There, in [4], the dipole
moment tensor $D^{ab}$ is decomposed into the electric dipole moment (EDM) $%
d^{a}$ and the magnetic dipole moment (MDM) $m^{a}$. It is also shown that
the spin four-tensor $S^{ab}$, which is an intrinsic angular momentum, can
be decomposed into two vectors, the usual `space-space'\ intrinsic angular
momentum $S^{a}$, which is called `magnetic' spin (mspin), and a new one,
the `time-space'\ intrinsic angular momentum $Z^{a}$, which is called
`electric' spin (espin). Both decompositions of $D^{ab}$ and $S^{ab}$ are
the same as in Minkowski's relations (\ref{M2}) and (\ref{M1}), i.e. as in
our equations (\ref{E2}) and (\ref{E1}). However, in the decompositions of $%
F $, $N$ and the angular momentum $M$ (bivector), in [23], the velocity
vector $v$ is the velocity vector of the observers, whereas in the
decompositions of $D^{ab}$ and the intrinsic angular momentum four-tensor $%
S^{ab}$ the velocity vector of the particle $u^{a}$ appears. Then, in [4],
the connection between $D^{ab}$ and the intrinsic angular momentum $S^{ab}$
is formulated in the form of the generalized Uhlenbeck-Goudsmit hypothesis, $%
D^{ab}=g_{S}S^{ab}$, equation (9) in [4]. Furthermore, equation (10) in [4],
it is proved that an MDM of a fundamental particle is determined by the
mspin $S^{a}$, as a vector correctly defined on the 4D spacetime, and not by
the usual 3D spin $\mathbf{S}$. Even more important result, again equation
(10) in [4], is that an EDM of a fundamental particle, as a vector, is
determined by the espin $Z^{a}$ and not, as generally accepted in the
standard model and in supersymmetric (SUSY) theories, by the usual 3D spin $%
\mathbf{S}$. In section 5 in [4] some shortcomings in the EDM searches are
discussed; they are all connected with the use of the UT instead of the LT.

In addition, it is obtained in [11, 12] that the usual formulation with the
3D $\mathbf{E}$ and $\mathbf{B}$ and their UT yields different values for
the emf for relatively moving inertial observers, see equations (27) and
(29) in [11] and equations (55) and (58) in [12]. On the other hand in the
approach with 4D geometric quantities the emf is defined as a Lorentz scalar
and consequently the same value for that emf is obtained for all relatively
moving inertial frames, see equations (35-37) in [11] and equations (61-63)
in [12].

\textbf{Conclusions}. - \emph{From the result that the transformations }(\ref%
{EP1}), (\ref{J1}) \emph{and }(\ref{J2}) \emph{are not the LT it can be
concluded that, contrary to the general opinion, neither the transformations}
(\ref{JC1}) and (\ref{JB}), \emph{i.e., the UT, equations }(11.148) \emph{and%
} (11.149) \emph{from} [2], \emph{are the LT.} Furthermore, the above
mentioned comparisons with experiments, the Trouton-Noble experiment [21,
22], the motional emf [11] and the Faraday disk [12] show that the approach
with multivectors always agrees with the principle of relativity and it is
in a true agreement (independent of the chosen inertial reference frame and
of the chosen system of coordinates in it) with experiments. As shown in
[11,12], [21, 22] and here, this is not the case with the usual approach,
e.g., [2], in which the electric and magnetic fields are represented by
3-vectors $\mathbf{E(r,}t\mathbf{)}$ and $\mathbf{B(r,}t\mathbf{)}$ that
transform according to the UT, or according to their two low-velocity limits
from [15,16].

Minkowski's great discovery of the correct LT (\ref{M3}), section 11.6 in
[8], their explicit forms (\ref{LM3})-(\ref{LTE2}) that are found in [10-13]
and also the mathematical argument from [9] that space and time dependent
electric and magnetic fields cannot be the usual 3-vectors strongly suggest
the need for further critical examination of the usual formulation of
electromagnetism with 3-vectors $\mathbf{E(r,}t\mathbf{)}$, $\mathbf{B(r,}t%
\mathbf{)}$ and their UT (\ref{JC1}) and (\ref{JB}) and also the possibility
for a complete and relativistically correct formulation of classical and
quantum electromagnetism with multivector fields (as physically real fields)
that are defined on the spacetime and with their correct LT. The advantages
of such formulation with multivector fields are already revealed in the
cases of the interaction between the dipole moment tensor $D^{ab}$ and the
electromagnetic field $F^{ab}$ in the first paper in [24] and in much more
detail in [4], in the discussion of quantum phase shifts in the second and
the third paper in [24] and in the formulation of Majorana form of the
Dirac-like equation for the free-photon [25].\bigskip \bigskip

\noindent \textbf{References\bigskip }

\noindent \lbrack 1] Einstein A 1905 \textit{Ann. Physik.} \textbf{17} 891
translated by Perrett W and Jeffery

G B 1952 in \textit{The Principle of Relativity} (New York: Dover)

\noindent \lbrack 2] Jackson J D 1998 \textit{Classical Electrodynamics} 3rd
edn (New York: Wiley)

\noindent \lbrack 3] Ivezi\'{c} T 2001 \textit{Found. Phys.} \textbf{31} 1139

\noindent \lbrack 4] Ivezi\'{c} T 2010 \textit{Phys. Scr. }\textbf{81} 025001

\noindent \lbrack 5] Lorentz H A 1904 \textit{Proceedings of the Royal
Netherlands Academy of Arts}

\textit{and Sciences} \textbf{6} 809

\noindent \lbrack 6] Poincar\'{e} H 1906 \textit{Rend. del Circ. Mat. di
Palermo} \textbf{21} 129

\noindent \lbrack 7] Logunov A A 1996 \textit{Hadronic J}. \textbf{19} 109

\noindent \lbrack 8] Minkowski H 1908 \textit{Nachr. Ges. Wiss. G\"{o}ttingen%
} 53 (The French translation

by Paul Langevin at http://hal.archives-ouvertes.fr/hal-00321285/fr/; the

Russian translation is in 1983 \textit{Einshteinovskii Sbornik} 1978-1979
(Moskva:

Nauka) pp 5-63)

\noindent \lbrack 9] Oziewicz Z 2008 \textit{Rev. Bull. Calcutta Math. Soc. }%
\textbf{16} 49

Oziewicz Z and Whitney C K 2008 \textit{Proc. Natural Philosophy }

\textit{Alliance (NPA)} \textbf{5} 183 (also at http://www.worldnpa.org/php/)

Oziewicz Z 2009 Unpublished results that can be obtained from the author

at oziewicz@unam.mx

\noindent \lbrack 10] Ivezi\'{c} T 2003 \textit{Found. Phys.} \textbf{33}
1339

\noindent \lbrack 11] Ivezi\'{c} T 2005 \textit{Found. Phys.} Lett. \textbf{%
18 }301

\noindent \lbrack 12] Ivezi\'{c} T 2005 \textit{Found. Phys.} \textbf{35}
1585

\noindent \lbrack 13] Ivezi\'{c} T 2008 \textit{Fizika A }\textbf{17} 1

\noindent \lbrack 14] Hestenes D 1966 \textit{Space-Time Algebra (}New York:
Gordon \& Breach)

Hestenes D and Sobczyk G 1984 \textit{Clifford Algebra to Geometric Calculus
}

(Dordrecht: Reidel)

Doran C. and Lasenby A 2003 \textit{Geometric algebra for physicists}

(Cambridge: Cambridge University Press)

\noindent \lbrack 15] Le Bellac M and L\'{e}vy-Leblond J M 1973 \textit{%
Nuovo Cimento B} \textbf{14} 217

\noindent \lbrack 16] Rousseaux G 2005 \textit{Europhys. Lett.} \textbf{71}
15

De Montigny M and Rousseaux G 2006 \textit{Eur. J. Phys.} \textbf{27} 755

De Montigny M and Rousseaux G 2007 \textit{Am. J. Phys.} \textbf{75} 984

Rousseaux G 2008 \textit{Phys. Rev. Lett.} \textbf{100} 248901

Rousseaux G 2008 \textit{Europhys. Lett.} \textbf{84} 20002

\noindent \lbrack 17] Ivezi\'{c} T 1999 \textit{Found. Phys.} \textit{Lett.}
\textbf{12} 507

\noindent \lbrack 18] Ivezi\'{c} T 2002 \textit{Found. Phys.} \textit{Lett.}
\textbf{15} 27

Ivezi\'{c} T 2001 \textit{arXiv: physics}/0103026

Ivezi\'{c} T 2001 \textit{arXiv: physics}/0101091

\noindent \lbrack 19] Ivezi\'{c} T 1999 \textit{Found. Phys.} \textit{Lett.,}
\textbf{12} 105

\noindent \lbrack 20] Jefimenko O D 1999 \textit{J. Phys. A: Math. Gen.}
\textbf{32} 3755

\noindent \lbrack 21] Ivezi\'{c} T 2005 \textit{Found. Phys.} \textit{Lett.}
\textbf{18} 401

\noindent \lbrack 22] Ivezi\'{c} T 2007 \textit{Found. Phys.} \textbf{37} 747

\noindent \lbrack 23] Ivezi\'{c} T 2006 \textit{Found. Phys.} \textbf{36}
1511

Ivezi\'{c} T 2007 \textit{Fizika A} \textbf{16} 207

\noindent \lbrack 24] Ivezi\'{c} T 2007 \textit{Phys. Rev. Lett.} \textbf{98}
108901

Ivezi\'{c} T 2007 \textit{Phys. Rev. Lett.} \textbf{98 }158901

Ivezi\'{c} T 2007 \textit{arXiv: hep-th}/0705.0744

\noindent \lbrack 25] Ivezi\'{c} T 2006 \textit{EJTP} \textbf{10} 131

\end{document}